\documentclass[pre,twocolumn,aps]{revtex4}

\usepackage{amssymb,amsmath,graphicx}

\begin{document}

\title{Clustering coefficient without degree correlations biases}

\author{Sara Nadiv Soffer$^1$ and Alexei V\'azquez$^2$}

\affiliation{$^1$ Department of Mathematics, Rutgers University
Piscataway, NJ 08854,  USA}

\affiliation{$^2$ Department of Physics and Center for Complex Network
Research, University of Notre Dame, IN 46556, USA}

\date{\today}

\begin{abstract}

The clustering coefficient quantifies how well connected are the neighbors
of a vertex in a graph. In real networks it decreases with the vertex
degree, which has been taken as a signature of the network hierarchical
structure. Here we show that this signature of hierarchical structure is a
consequence of degree correlation biases in the clustering coefficient
definition. We introduce a new definition in which the degree correlation
biases are filtered out, and provide evidence that in real networks the
clustering coefficient is constant or decays logarithmically with vertex
degree.

\end{abstract}

\maketitle

\bibliographystyle{apsrev}

\begin{figure}[t]
\centerline{\includegraphics[height=1in]{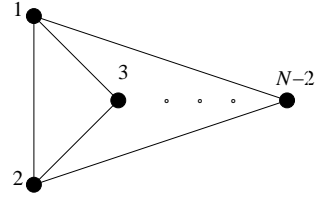}}

\caption{Double star with two vertices, 1 and 2, connected to $N-2$ other
vertices. The neighbors of vertex 1 (or 2) are connected as most as their
degrees allow. Yet, with the old definition of clustering coefficient
we obtain $c_1={\cal O}(1/N)$, approaching zero in the limit $N\gg1$.}

\label{fig5}
\end{figure}

The increasing availability of network data representing many real systems
have motivated the development of new statistical measures to characterize
large networks \cite{ab01a,dm002b,newman03a,hgn,pv04}. These measures has
revealed that, as a difference with the classical Erd{\"o}s R\'{e}nyi
\cite{er59} random graph model, real networks are characterized by a power
law distribution of vertex degrees \cite{fff99,bajb00a,ab01a}, a high
clustering coefficient or transitivity \cite{ws98,ab01a}, and degree
correlations between connected vertices \cite{pvv01,n02a,ms02}. Yet, it is
important to characterize up to which extent the new measures provides new
information about the studied networks. For instance, it has been shown
that in some networks the degree correlations are a consequence of the
existence of large degree vertices and, therefore, the sequence of vertex
degrees is sufficient to characterize those networks
\cite{ms02,park03,catanzaro04}.

In this work we study the influence of degree correlations on the
clustering coefficient.  We show that most of the observed variations of
the clustering coefficient with the vertex degrees
\cite{vpv02a,ravasz02,ravasz03,vazquez03} are determined by the degree
correlations among connected vertices. Based on this fact, we introduce a
new definition of clustering coefficient, filtering out the effect of
degree correlations. The similarities and differences between the two
definitions are analyzed through the study of different real networks.

Consider undirected simple graphs on $i=1,\ldots,N$ vertices. Let
$k_i$ be the degree of a vertex and $t_i$ the number of edges
among its neighbors. The standard definition of local clustering
coefficient is

\begin{equation}
c_i = \frac{t_i}{\binom{k_i}{2}}\ , \label{ci}
\end{equation}

\noindent where $\binom{k_i}{2}$ is the number of pairs that can be made
using $k_i$ neighbors. Furthermore, to characterize the global clustering
coefficient two different measures has been introduced. The first is just
the average of $c_i$ over all vertices with degree larger than one

\begin{equation}
\left<c\right> = \frac{\sum_{i|k_i>1}c_i }{ \sum_{i|k_i>1} 1}\ .
\label{cave}
\end{equation}

\noindent The second is obtained computing first the average of $t_i$ and
$\binom{k_i}{2}$ and then their ratio

\begin{equation}
C = \frac{\sum_i t_i}{\sum_i \binom{k_i}{2}}\ .
\label{C}
\end{equation}

\noindent As noticed in \cite{bollobas03}, the two definitions of global
clustering coefficient may give different values. Consider, for instance,
a double star of $N$ vertices (Fig. \ref{fig5}). In this case
$\left<c\right>\approx1$, $C={\cal O}(1/N)$ and the two global clustering
coefficients dramatically differ for $N\gg1$. The limitations in the
clustering coefficient definition are not only related to the way averages
are computed. The local clustering coefficient of any of the two central
vertices of the double star, vertex 1 for instance, is $c_1={\cal
O}(1/N)$, approaching zero for $N\gg1$. We cannot, however, increase the
number of connections among the neighbors of vertex 1 without increasing
the degree of its neighbors. In this sense, the neighbors of vertex 1 are
as clustered as they can be, in contradiction with the small value of
$c_1$.

\begin{figure}[t]
\centerline{\includegraphics[width=3.3in]{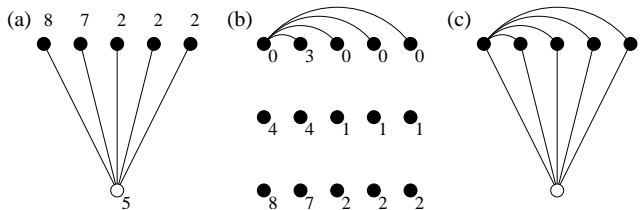}}

\caption{Algorithm to compute $\omega_i$. (a) A vertex $i$ (open
circle) is connected to five neighbors (filled circles) with
degree sequence $\left\{8,7,2,2,2\right\}$. (b) Since each
neighbor can be connected at most with four other neighbors, we
replace the neighbors degree sequence (lowest raw) by
$\left\{4,4,1,1,1\right\}$ (middle raw).  It is easy to see that
after connecting the first neighbor to all others, we get 4
triangles and 3 extra edges that can't be used anymore (upper
raw). Summarizing, for this example, $\omega_i = 4$, $\Omega_i=5$
and $\binom{5}{2}=10$. (c) Subgraph with maximum number of edges
among the neighbors, with $c_i=0.4$ and $\tilde{c}_i=1$.}

\label{fig4}
\end{figure}

This example shows that the local clustering coefficient of a large degree
vertex connected to vertices with much smaller degrees will be always
small, no matter how its neighbors are interconnected. We would like
instead a measure of clustering coefficient that allow us to quantify the
connectivity among the neighbors of a vertex, independently of its degree
and the degree of its neighbors. The clustering coefficient is a three
vertex correlation measure and, as it is the general case in statistics,
to define a three point correlation measure we should filter out two point
correlations, represented here by the degree correlations between
connected vertices. We tackle this problem defining the clustering
coefficient relative to the maximum possible number of edges between the
neighbors of a vertex, given their degree sequence.  Let $\omega_i$ be the
maximum number of edges that can be drawn among the $k_i$ neighbors of a
vertex $i$, given the degree sequence of its neighbors. A neighbor $j$ can
have at most $\min(k_i-1,k_j-1)$ edges with the other neighbors, therefore

\begin{equation}
\omega_i\leq \Omega_i = \left\lfloor\frac{1}{2}\sum_{\mbox{neighbors}}
\left[ \min(k_i,k_j)- 1 \right]\right\rfloor \leq
\binom{k_i}{2}\ . \label{omegai}
\end{equation}

\noindent While $\binom{k_i}{2}$ takes into account only the degree of the
vertex, $\Omega_i$ considers that occasionally, not all the $k_i-1$ excess
edges are available at the neighbors of $i$. $\omega_i$ considers, in
addition, the possibility of the excess edges to actually form triangles.
$\omega_i$ can be computed using the following algorithm \cite{soffer04b}:
1- Starting from the neighbor's degree sequence $\{k_1,\ldots,k_n\}$
($n=k_i$), construct the list $\{\min(k_i,k_1)-1, \ldots,
\min(k_i,k_n)-1\}$, arranged in a decreasing order. 2- Draw an edge from
the first element to as many as possible other elements in the list,
always going from largest to smaller. Each time an edge is drawn, one is
subtracted from the remaining degree of the connected vertices. 3 - Remove
the first element and any zero from the list and sort the list in
decreasing order. 4- Repeat the process and stop when the list is empty.
The number of maximum possible connections $\omega_i$ is the total number
of edges drawn (see Fig. \ref{fig4}).

A proper definition of local clustering coefficient,
removing the effects of degree correlations, is

\begin{equation}
\tilde{c}_i = \frac{t_i}{\omega_i}\ . \label{c1i}
\end{equation}

\noindent and the two different measures of global clustering coefficient
read

\begin{equation}
\left<\tilde{c}\right> = \frac{\sum_{i|\omega_i>0}c_i}
{\sum_{i|\omega_i>0}1}\ ,\ \ \ \ \ \ \
\tilde{C} = \frac{\sum_i t_i}{\sum_i \omega_i} \ .
\label{c1aveC1}
\end{equation}

\begin{table}[t]
\begin{tabular}{|l|l|l|l|l|l|}
\hline
Network & $r$ & $<c>$ & $C$ & $<\tilde{c}>$ & $\tilde{C}$\\
\hline
Internet & -0.19 & 0.45 & 0.0090 & 0.49 & 0.45\\
protein interaction & -0.13 & 0.12 & 0.055 & 0.16 & 0.19\\
semantic & 0.085 & 0.75 & 0.31 & 0.83 & 0.59\\
co-authorship & 0.67 & 0.65 & 0.56 & 0.78 & 0.85\\
\hline
\end{tabular}

\caption{Average clustering coefficient as computed with the old
and new definitions. The graphs are listed in increasing order
of their degree of assortativity, quantified by the degree
correlation coefficient $r$ \cite{n02a}, taking values from -1 (fully
disassortative) to 1 (fully assortative).}

\label{tab1}
\end{table}

\noindent Some general properties of the new definition of clustering
coefficient are the following. ({\it i}) If all the neighbors of a vertex
has degree one (star) then its clustering coefficient is undefined.
Indeed, the concept of clustering is meaningless for the central vertex of
a star, as it is meaningless for degree one vertices. ({\it ii})
$\tilde{c}_i\geq c_i$, as follows from (\ref{omegai}). Therefore, when the
clustering is one by the old definition it is one by the new definition.
Notice that the opposite is not necessarily true (see Fig. \ref{fig4}(c)).
({\it iii}) When all the $k_i$ neighbors of a vertex $i$ have degrees
larger or equal to the degree of the vertex itself (a regular graph, for
instance) $\tilde{c}_i=c_i$.

The example in Fig. \ref{fig4} shows how the old definition
underestimates the clustering around a given vertex $i$. In this
case, the number of edges between neighbors is as large as it can
be given their degree sequence. This picture is not captured by
the clustering coefficient according to the old definition
($c_i=0.4$), but it is correctly quantified using the new
definition ($\tilde{c}_i=1$). In the following we compare the old
and new clustering coefficient definitions using the graph
representation of four real systems. The degree of correlations
present on these graphs is quantified by the assortativity
coefficient $r$ \cite{n02a}, taking values between -1 (highly
disassortative) to 1 (highly assortative). The systems considered
are, in increasing order of assortativity, 1- the autonomous
system representation of the Internet, as for April 2001
\cite{nlanr}, 2- the protein-protein interaction network of the
yeast {\it Saccharomyces cerevisiae} \cite{dip}, 3- the semantic
web of English synonyms \cite{ravasz03}, and 4- the co-authorship
network of mathematical publications between 1991 and 1999
\cite{bjnr01a}. In Table \ref{tab1} we show the two global
clustering coefficients as computed with the old and new
definitions.  For the two disassortative graphs ($r<0$), there is
an orders of magnitude difference between the global clustering
coefficients $\left<c\right>$ and $C$ computed with the old
definition. With the new definition, however, both global measures
of clustering coefficient (\ref{c1aveC1}) gives values of the same
order, independently of the degree correlations.

\begin{figure}
\centerline{\includegraphics[width=3in,height=8in]{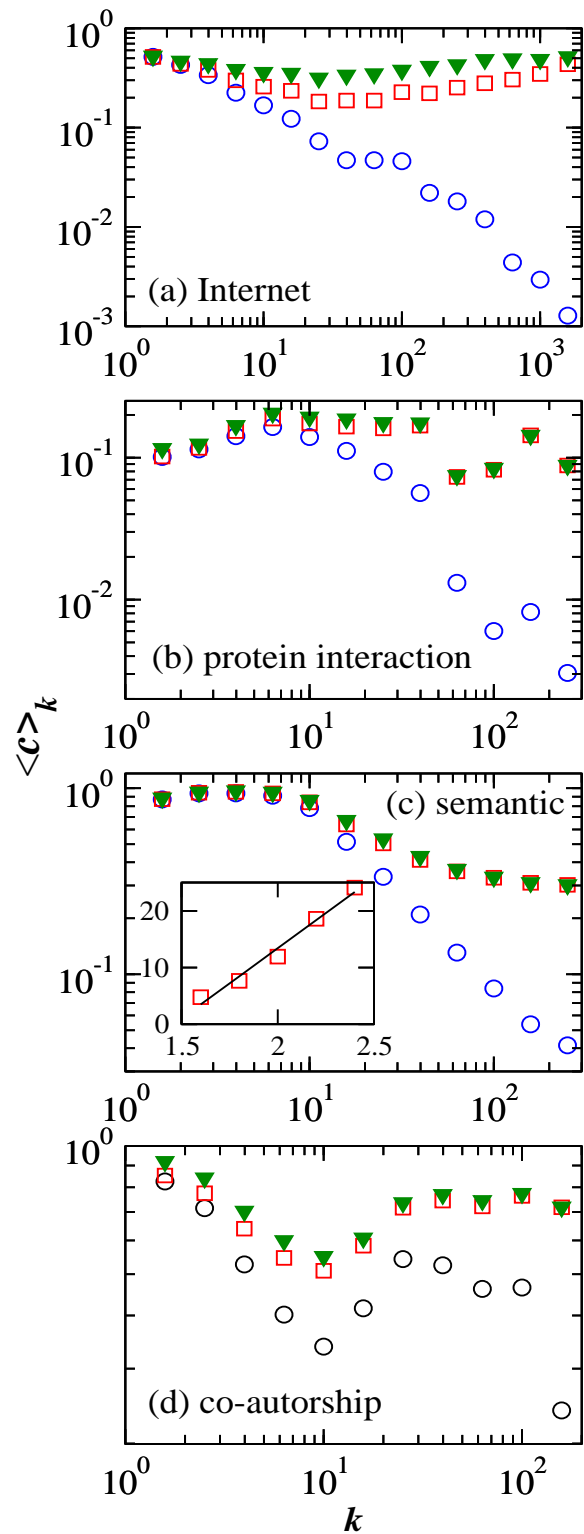}}

\caption{Average clustering as a function of the vertex degree, as
computed using the old definition (circles), the new definition
approximating $\omega_i$ by $\Omega_i$ (squares) and the new definition
using $\omega_i$ (triangles). The graphs are shown in increasing order of
their assortativity, with the most disassortative graph in the top, and
the more assortative graph on the bottom.}

\label{fig2}
\end{figure}

Another characteristic feature of the old definition of clustering
coefficient is that, when the average is restricted to vertices
with the same degree $\left<c\right>_k$, it decays as
$\left<c\right>_k\sim k^{-\alpha}$ with vertex degree
\cite{vpv02a,ravasz02,ravasz03,vazquez03}. This decay can be
observed in Fig. \ref{fig2} for the four graphs considered here,
being more pronounced for the two disassortative graphs in Fig.
\ref{fig2}(a) and (b), and almost absent for the highly
assortative co-authorship graph in Fig. \ref{fig2}(d).  In
contrast, when computed with the new definition (\ref{c1i}),
$\left<\tilde{c}\right>_k$ does not exhibit a strong variation
with increasing vertex degree (see Fig. \ref{fig2}).

In particular, the decreasing trend is completely absent for 
the Internet (Fig. \ref{fig2}(a)), and the variations between the
smallest and largest new clustering coefficient are no more than a factor
of two, indicating variations previously observed with the standard
definition \cite{vpv02a} are reflecting degree correlations.  The large
variations of $\left<c\right>_k$ with the vertex degree $k$ have been
interpreted as the existence of a hierarchical structure, with high degree
vertices interconnecting highly connected subgraphs made of smaller degree
vertices, but with no or few connections among vertices in different
subgraphs \cite{vpv02a,ravasz02}. The existence of this hierarchical
structure, however, was already predicted from the analysis of the degree
correlations \cite{pvv01,pv04}. The present work make the bridge between
these two differences approaches to quantify the hierarchical structure of
the Internet, showing that the variations in the clustering coefficient
with the vertex degrees, as measured with the old definition, are just
reflecting the existence of degree correlations. These conclusions are
also applicable for the protein-protein interaction graph, with a degree
of disassortative close to that of the Internet graph.

In the case of the Internet we can also follow changes in the clustering
coefficient as the network evolves, with around 3000 vertices in 1997 to
10000 vertices in 2001. $\left<\tilde c\right>_k$ remains essentially
stationary within this period (data not shown), as does $\left<c\right>_k$
\cite{vpv02a}. In contrast, in random graphs with fixed degree
distribution and degree correlations the local clustering coefficient
approaches zero with increasing graph size, independently of the vertex
degree \cite{dorogovtsev03a}. Therefore, the Internet is more clustered
than expected from the degree distribution and degree correlations alone.

In the case of the semantic web (Fig. \ref{fig2}(c)), although the
clustering coefficient variations are reduced after filtering out the
degree correlations, there is still a logarithmic decrease with increasing
the vertex degree (see inset of Fig. \ref{fig2}(c)). Using a deterministic
growing graph model introduced in \cite{dgm02a}, we show that this
logarithmic decay may be the general case for graphs where
$\left<c\right>_k\sim 1/k$. In the deterministic model, we start with one
edge at time $t=-1$. At each time step we create a new triangle on each
existing edge by connecting its two endpoints to a new vertex. At time
$t=0$ we get one triangle and at time $t=1$, we will have the triangle
from the previous step and three new ones, each is using one edge from the
existing old triangle and two new edges with a new vertex between. Since
this model is built recursively, we can find by induction the degree of a
vertex $k_i(\tau) = 2^{\tau+1}$ and the number of triangles that passing
trough it $t_i=k_i-1$, where $\tau$ is the time elapsed from the
introduction of the vertex, resulting the clustering coefficient
$c_i=2/k_i$ \cite{dgm02a}. To compute the clustering coefficient according
to the new definition (\ref{c1i}) we need to determine the scaling of
$\omega_i$ with the vertex degree $k_i$. From the $\Omega_i$ definition
(\ref{omegai}) and the evolution rules of the model we obtain the
following recursive relation $\Omega_i(\tau + 1) = 2\Omega_i(\tau)+2^{\tau
+ 1}$. From this recursive relation and the initial condition $\Omega_i(0)
= 1$ we obtain by induction $\Omega_i( \tau) = ( \tau + 1 ) 2^{\tau}$. We
have also obtained an exact expression for $\omega_i$ \cite{soffer04b},
which in the $\tau\gg1$ limit results in $\omega_i\approx\Omega_i$ and

\begin{equation}
\tilde{c}_i \approx \frac{2}{\log_2{k_i}} \ , \label{cnewapp}
\end{equation}

\noindent The analysis of the deterministic model indicates that in graphs
where the old definition of clustering coefficient is characterized by an
inverse proportionality with the vertex degree, the new clustering
coefficient will exhibit a logarithmic decrease with increasing the vertex
degree. This observation is in agreement with the semantic web data as
well (Fig.\ref{fig2}c), where $\left<c\right>_k\sim1/k$ and
$\left<\tilde{c}\right>_k\sim1/\log k$.

Finally, for the most assortative graph in Fig. \ref{fig2}(d), we do not
observe a substantial difference between the two definitions of clustering
coefficient. This observation is explained by the fact that in a highly
assortative graph the degree of connected vertices is quite similar,
$\omega_i\approx\Omega_i\approx\binom{k_i}{2}$ and the two clustering
coefficient definitions give similar results.

We have shown evidence that the standard definition of clustering
coefficient in Eq. (\ref{ci}) contains some biases due to the
degree correlations between connected vertices.  After removing
these biases the local clustering coefficient does not depend
strongly on the vertex degrees, being of the same order for small
and large degree vertices. More precisely, we observe two
different scenarios, either the local clustering coefficient is
approximately constant or it decays logarithmically with
increasing the vertex degree. These results will eventually force
us to reevaluate the clustering based analysis of complex
networks, and other approaches
\cite{ravasz02,newman03b,barrat04,vazquez04} based on this
magnitude.

We thank A.-L. Barab\'asi and A. Vespignani for helpful comments and
discussion.


\end{document}